\documentclass{article}
\usepackage{graphicx} % Required for inserting images
\usepackage{amsmath}
\usepackage{amsfonts}
\usepackage{amssymb}
\date{\today}
\usepackage{graphicx} % This package allows for the inclusion of images
\usepackage{float}

\newcommand{\insertplot}[5]{\begin{figure}
 \hfill\hbox to 0.05in{\vbox to #5in{\vfill
 \inputplot{#1}{#4}{#5}}\hfill}
 \hfill\vspace{-.1in}
 \caption{#2}\label{#3}
 \end{figure}}
 \newcommand{\inputplot}[3]{
 \special{ps: plotfile #1}
}
\newcounter{fig}

\newcommand{\ee}{\end{equation}}
\newcommand{\eea}{\end{eqnarray}}
\newcommand{\be}{\begin{equation}}

\begin{document}

\title{Hairy charged black holes \\
in a model with bounded scalar field potential}

\date{\today}

\author{
{\large Yves Brihaye}$^{1}$,
{\large Betti Hartmann} $^{2}$, and
{\large Katherine Horton}$^{2}$
\\
\\
$^{1}${\small Physique de l'Univers, Universit\'e de
Mons, 7000 Mons, Belgium}
\\
$^{2}${\small Department of Mathematics, University College London, Gower Street, London, WC1E 6BT, UK}
}
%%%
\maketitle

\begin{abstract}
Electrically and magnetically charged (a.k.a. dyonic) black holes with scalar hair have recently been constructed for a sextic scalar field potential. Here, we re-investigate this model, but with a bounded scalar field
potential of exponential form. We demonstrate that qualitative differences appear. 
First, we present scalar clouds on (dyonic) Reissner-Nordstr\"om black holes as well as (dyonically) charged clouds on Schwarzschild black holes, respectively. We then extend our results to the fully backreacted case and put the focus on the comparison of the electrically charged black holes with their dyonic counterparts.  

\end{abstract}

%%%%%%%%%%%%%%%%%%%%%%%%%
\section{Introduction}
%%%%%%%%%%%%%%%%%%%%%%%%

The no-scalar hair theorems \cite{noscalarhair} put restrictions on the existence of scalar hair on the horizon of a black hole. In an attempt to understand how high energy physics works in curved space-time, particle physics models involving scalar fields have been consider and it was shown that black holes with non-trivial scalar hair can exist. Examples of these 
are black holes that carry Skyrme hair \cite{Luckock:1986tr} as well as Higgs hair \cite{Breitenlohner:1991aa}. In both of these cases, the flat space-time limit possesses topological soliton solutions, in the former case the Skyrmion \cite{Skyrme:1961vq}, in the latter
the 't Hooft-Polyakov magnetic monopole \cite{tHooft:1974kcl,polyakov}. 
Black holes with hair can then be thought of as residing in the center of these solitonic objects. Note that the SU(2) Yang-Mills-Higgs model does not only allow
black holes inside monopoles, but also magnetically charged Reissner-Nordstr\"om solutions. The unbroken U(1) gauge field is associated to the long-range magnetic field. Boson stars are gravitating non-topological solitons \cite{boson_stars, boson_stars_rot} that are made of a complex valued scalar field  and possess a globally conserved Noether charge. That black holes cannot be put into
the center of uncharged and non-rotating boson stars is due to the fact that the radial pressure associated to the scalar field is larger than the pressure in angular direction \cite{Pena:1997cy}. However, when either giving the solution
angular momentum \cite{scalar_Kerr} and/or gauging the U(1) symmetry of the complex scalar field model \cite{Hong:2020miv,Herdeiro:2020xmb,Brihaye:2020vce} , black hole solutions with scalar hair can be constructed. 
A synchronisation condition has to be fulfilled which assigns either the horizon velocity or the electric potential at the
horizon  to the frequency of the complex scalar field, respectively.
This condition appears exactly at the threshold of superradiance. 

Recently, new types of spherically symmetric, static black holes which carry scalar hair were constructed \cite{Herdeiro:2024yqa}. These solutions possess both electric and magnetic fields with the magnetic field representing that of a magnetic monopole sitting at the origin of the coordinate system behind the event horizon. When magnetic fields are present, a complex doublet of scalar fields is necessary. This possesses a 
global $SU(2)$ symmetry and the dependence on the coordinates is such that 
the field equations remain cohomogeneity-one. 
The minimal coupling  of the scalar field  to the magnetic monopole charge requires the Dirac quantisation condition \cite{dirac}. 
Accordingly, there exists no regular limit of the corresponding black hole solutions.  

In this paper we follow the investigation of \cite{Herdeiro:2024yqa}, but
now choose a scalar field potential that is bounded. We discuss in detail the case of uncharged scalar fields in the background of an electrically and magnetically charged Reissner-Nordstr\"om black hole as well as electrically and magnetically charged scalar fields in the background of a Schwarzschild black hole, respectively.
These limiting cases have not been investigated in detail before. 
We also discuss the electrically and magnetically charged black holes with scalar hair and compare our results to those obtained in 
\cite{Herdeiro:2024yqa}, where a polynomial self-interacting scalar field potential was used.

We discuss the model in Section 2, scalar clouds on black holes in Section 3 and black holes with scalar hair in Section 4, respectively. We conclude in Section 5.

%%%%%%%%%%%%%%%%%%%%%%
\section{The model}
%%%%%%%%%%%%%%%%%%%%%
The action of the model we are studying here reads:
\begin{equation}
\label{eq:action}
{\cal S} = \int {\rm d}^4 x \sqrt{-g} \left(\frac{{\cal R}}{4\alpha} + {\cal L}_{m}\right)
\end{equation}
where $\alpha=4\pi G$. This is the Einstein-Hilbert action with Ricci scalar ${\cal R}$ and Newton's constant $G$ minimally coupled to matter fields with Lagrangian density ${\cal L}_m$. For the energy-momentum content
we consider  a number $n+1$ of self-interacting, massive and complex scalar fields $\Phi_k$, $k=1,...,n+1$ 
minimally coupled to  a U(1) gauge field $A_{\mu}$  with Lagrangian density
\begin{equation}
\label{eq:matter_lagrangian}
{\cal L}_{m}= \sum\limits_{k=1}^{n+1} \left[-D_{\mu} \Phi_k (D^{\mu}\Phi_k)^{*} -U(\Phi_k\Phi_k^*)\right] -\frac{1}{4} F_{\mu\nu} F^{\mu\nu} \ \ ,
\end{equation}
where $D_{\mu} = \partial_{\mu} - i g A_{\mu}$ is the covariant derivative with gauge coupling $g$ and  $F_{\mu\nu}=\partial_{\mu}A_{\nu} - \partial_{\nu}A_{\mu}$ is the U(1) field strength tensor. 
In \cite{Herdeiro:2024yqa} this model was studied with a sextic potential, here we will  use an exponential potential of the form~:
\be
\label{eq:susy_pot}
             U(\Phi_k\Phi_k^*) = m^2 \eta^2 \Bigr(1 - \exp(- \Phi_k\Phi_k^* / \eta^2) \Bigl) 
\ee
where $\eta$ is an energy scale. Note that unlike the sextic potential used in \cite{Herdeiro:2024yqa} this potential is bounded in $|\Phi|$.

%%%%%%%%%%%%%%%%%%%%%%%%%%%%%%%%%%%%%%%%%%%%%%%%%
\subsection{The Ansatz and equations of motion}
%%%%%%%%%%%%%%%%%%%%%%%%%%%%%%%%%%%%%%%%%%%%%%%%%
In the following, we will study spherically symmetric solutions and hence use the Ansatz
\be
\label{eq:metric}
     {\rm d}s^2 = - N(r) \sigma^2(r) {\rm d}t^2 + \frac{1}{N(r)} {\rm d}r^2 + r^2 {\rm d}\theta^2 + r^2 \sin^2\theta {\rm d}\varphi^2  \ \ , \ \  N(r)=1-\frac{2\mu(r)}{r}
%%%%		\ \ 
\ee
for the metric with $\mu(r)$ the mass function and \cite{Herdeiro:2024yqa}
\be
\label{electromagnetic}
               A_{\mu} {\rm d}x^{\mu} = V(r) {\rm d}t + Q_m \cos (\theta) {\rm d} \varphi
\ee
for the U(1) gauge field. The parameter $Q_m$ denotes the magnetic charge of the solution. For the scalar field, we distinguish the cases $k=0$ and $k=1$, where $k \in {\mathbb{N}}$ appears in the Dirac quantisation condition ($c=\hbar=1$):
\be
                  g Q_m = \pm \frac{k}{2} \ \ , \ \ k \in \mathbb{N} \ .
\ee
The electric and magnetic fields of the solutions is given by
\begin{equation}
E_r=-F_{r0}=-\frac{{\rm d} V}{{\rm d}r} \ \ , \ \   B_r=\frac{F^{\theta\varphi}}{\sqrt{-g}}  = -\frac{Q_m}{r^2} \ ,
\end{equation}
where the latter describes the field of a magnetic monopole located at the origin. 
In the following, the choice of Ansatz for the scalar field depends on whether we choose $Q_m=0$ or $Q_m\neq 0$. For $k=0$, i.e. scalar fields charged only electrically, we choose
\be
 \Phi = \phi(r) e^{-i\omega t} 
 \ee
 and for $k=1$, i.e. electrically and magnetically charged scalar fields, we choose \cite{Herdeiro:2024yqa}
 \be
\Phi_1 = \phi(r)  \sin\left(\frac{\theta}{2}\right) e^{i( \varphi/2 -\omega t)}\ \  \ , \ \ \  
		\Phi_2 = \phi(r)  \cos\left(\frac{\theta}{2}\right) e^{-i( \varphi/2 +\omega t)}
\ee
The equations of motion that result from the variation of (\ref{eq:action}) are invariant under the following rescaling of the fields and coupling constants
\be
  \phi \rightarrow \frac{\phi}{\eta} \ \ , \ \ r \rightarrow m r \ \ , \ \ V \rightarrow \frac{V}{\eta} \ \ , \ \ 
	 Q_m \rightarrow \frac{m}{\eta} Q_m \ \ , \ \ g \rightarrow \frac{\eta}{m} g \ \ ,
\ee
which, in particular, leaves the Dirac quantization invariant. We will use these
rescalings to set the scalar field mass $m=1$, $\eta=1$ without loss of generality. With this choice, the field equations read~: 
\begin{eqnarray}
\mu' &=& \alpha \left[r^2 N \phi'^2 + \frac{r^2 g^2 V^2 \phi^2}{\sigma^2 N} + \frac{k}{2}\phi^2 + \frac{r^2 V'^2}{2\sigma^2} + \frac{k^2}{8g^2 r^2} + r^2 \left(1-\exp(-\phi^2)\right)\right] \ , \\
\frac{\sigma'}{\sigma} &=& 2\alpha r\left[\phi'^2 + \frac{g^2 V^2 \phi^2}{\sigma^2 N^2}\right] \ , \\
\left(\frac{r^2 V'}{\sigma}\right)' &=& 2 \frac{r^2 g^2 V \phi^2}{\sigma N} \ , \\
\left(\sigma r^2 N \phi'\right)' &=& -\frac{r^2 g^2 V^2 \phi}{\sigma N} + \frac{k}{2} \sigma \phi + r^2 \sigma \phi \exp(-\phi^2) \  .
\label{eq:phi}  
\end{eqnarray}
Note that  the synchronisation condition for black holes with scalar hair reads $\omega = g V_{h}$, 
where $V_{h}=V(r_h)$ is the value of the electric potential on the horizon $r_h$ and we have chosen $\omega=V_{h}=0$ in the equations above. Also note that the effective mass of the scalar field is given by
\begin{equation}
\label{eq:meff}
m_{\rm eff}^2 := m^2 - g^2 V_{\infty}^2 
\end{equation}
where $V_{\infty}$ is the value of the gauge field $V(r)$ at $r\rightarrow \infty$. We also require $m_{\rm eff}^2 \geq 0$ which ensures that the scalar field is exponentially localised. Since $V(r_h)=0$,
the value of $V_{\infty}$ determines the potential difference between the horizon and infinity.
Requiring $m^2 \geq  g^2 V_{\infty}^2$ then means that we cannot choose $g$ and/or $V_{\infty}$ too large, otherwise, we would be able to create scalar particles of mass $m$.

The model possesses a U(1) symmetry that leads to the conserved Noether charge
\begin{equation}
Q_{N}= \int\limits_{r_h}^{\infty} {\rm d} r \ r^2 \frac{2g V \phi^2}{N\sigma} \ .
\end{equation}
The Noether charge $Q_{N}$ can be thought of as the number of scalar bosons making up the cloud that surrounds the black hole horizon. Due to the coupling to the electromagnetic field, these scalar bosons possess each an electric charge $g$, such that the electric charge in the scalar cloud is $gQ_{N}$. 
When the matter fields are not coupled to the gravitational fields, we define the mass $M_Q$ of the cloud as 
\begin{equation}
M_Q= \frac{1}{4\pi} \int {\rm d}^3 x \ \sqrt{-g} \left( T^i_i-T^0_0\right) \ .
\end{equation}
In the following, we will also use the ADM mass $M_{\rm ADM}$ for the fully back-reacted case, where
$M_{\rm ADM}=\mu(r\rightarrow \infty)$. Finally, the Hawking temperature of the black holes reads
\begin{equation}
T_H=\frac{1}{4\pi} \sigma(r_h) N^{\prime}(r)\vert_{r=r_h}  \ .
\end{equation}

%%%%%%%%%%%%%%%%%%%%%%%
\section{Scalar clouds on black holes}
%%%%%%%%%%%%%%%%%%%%%%%%%%%%%%%%%
The scalar field not back-reacting on the space-time is an interesting limit that has not been investigated in detail in \cite{Herdeiro:2024yqa} or anywhere else in the literature. We will close this gap here and
discuss two cases separately: (a) the scalar field equation
decoupled from the Einstein-Maxwell equations,  and (b)  both matter fields (scalar and electromagnetic)
decoupled from gravity. In the former case this leads to the study of scalar fields in the background of (dyonic) Reissner-Nordstr\"om solutions, while in the latter case, the background is the Schwarzschild solution and the scalar field is electrically (and magnetically) charged. 

%%%%%%%%%%%%%%%%%%%%%%%%%%%%%%%%%%%%%%%%%%%%%%%%%%%%%%%%%%%%%%%%%%%%%%%%
\subsection{Scalar clouds on (dyonic) Reissner-Nordstr\"om black holes}
%%%%%%%%%%%%%%%%%%%%%%%%%%%%%%%%%%%%%%%%%%%%%%%%%%%%%%%%%%%%%%%%%%%%%%%%
In order to decouple the scalar field from the Einstein-Maxwell equations
we rescale $V \rightarrow V/\sqrt{\alpha}$ and $g\rightarrow \sqrt{\alpha} g$ and let $\alpha=0$. The solution to the Einstein-Maxwell equations is the electrically (and magnetically) charged 
Reissner-Nordstr\"om (RN) solution with metric functions~:
\be
\label{eq:solRNN}
  \sigma(r)\equiv 1 \ \ \ , \ \ \     N = 1 - \frac{2M_{\rm ADM}}{r} +  \frac{\left(Q_e^2 + Q_m^2\right)}{r^2}  
       = \frac{(r-r_+)(r-r_-)}{r^2}
\ee
with the two horizons $r_{\pm}=2M_{\rm ADM} \pm \sqrt{M_{\rm ADM}^2-(Q_e^2 + Q_m^2)}$. $Q_e$ denotes the electric charge, while $Q_m=\pm k/(2g)$ is the magnetic charge. The mass $M_{\rm ADM}$ can be written in terms of the event horizon radius $r_+=r_h$ and the charges as follows
\begin{equation} M_{\rm ADM}=\frac{r_h}{2} \left(1 + \frac{\left(Q_e^2 + \frac{k^2}{4g^2}\right)}{r_h^2}\right) \ \ . \ 
\end{equation}
The electric potential for this solution is then
\be
\label{eq:solRNV}
			 V(r) =  Q_e \left(\frac{1}{r_h} - \frac{1}{r}\right) =  V_{\infty} -\frac{Q_e}{r}\ \ 
\ee
where we have defined the value of the electric potential at infinity $V_{\infty} := Q_e/r_h$. The remaining equation to be solved
is the scalar field equation (\ref{eq:phi}) subject to appropriate boundary conditions.
Requiring regular scalar fields on the horizon $r=r_h$ leads to~:
\be
\label{eq:bc1}
          (N'\phi')\left|_{r=r_h} \right. = \frac{1}{2} \frac{dU}{d \phi}\left|_{r=r_h} + \frac{k \phi(r_h)}{2 r_h^2} \right. \ \ .
\ee
For the effective mass (\ref{eq:meff}) we find in this case 
\be
\label{eq:conditionex}
          m_{\rm eff}^2 =  m^2 -\frac{g^2 Q_e^2}{r_h^2} \  .
\ee
This leads to an upper bound on the value of $g$. Moreover, it is well known that extremal black holes cannot carry scalar hair. Hence, 
we also need to require that $M_{\rm ADM}^2 > Q_e^2 +Q_m^2$, which leads to a lower bound on $g$.
Combining these two bounds, we find~:
\be
\label{eq:g_condition}
                \frac{k}{2}  \sqrt{\frac{1}{r_h^2 - Q_e^2}} <    g < \frac{m r_h}{Q_e} \ .
\ee
Note that in the following, as stated above, $m=1$ without loss of generality. 

%%%%%%%%%%%%%%%%%%%%%%%%%%%%%%%%%%%%%%%%%%%%%%%%%%%%%%%%%%%
\subsubsection{Scalar clouds on electrically charged RN black holes}
%%%%%%%%%%%%%%%%%%%%%%%%%%%%%%%%%%%%%%%%%%%%%%%%%%%%%%%%%%%%
We first discuss the case $k=0$, i.e. a scalar field in the background of an electrically charged RN black hole. The scalar field equation (\ref{eq:phi}) needs to
be solved numerically. Fixing $r_h=1.0$ and $Q_e$ to different values, our numerical results strongly suggest that scalar clouds exist for $g\in  g \in \  ] 0, g_{\rm max}]$ with  $g_{\rm max} = \frac{m r_h}{Q_{e}}$ in agreement with (\ref{eq:g_condition}). In the limit $g\rightarrow g_{\rm max}$ we find that both $\phi(r_h)\rightarrow 0$ and $m_{\rm eff}^2 \rightarrow 0$, i.e. the scalar field function becomes equivalent to the null function.  For  $g$ decreasing, on the other hand, the value of the scalar field on the horizon, $\phi(r_h)$, increases. 
The dependence of $gQ_e$ and $m_{\rm eff}^2$ on the parameter $\phi(r_h)$ is qualitatively and quantitatively practically unchanged when considering different values of $Q_e$.

%%%%%%%%%%%%%%%%%%%%%%%%%%%%%%%%%%%%%%%%%%%%
\subsubsection{Scalar clouds on electrically and magnetically charged RN black holes}
\begin{figure}[h!]
\begin{center}
\includegraphics[width=8cm]{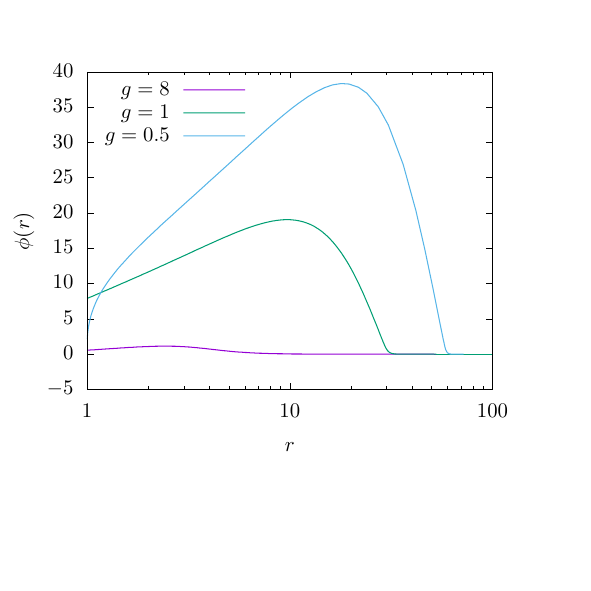}
\includegraphics[width=8cm]{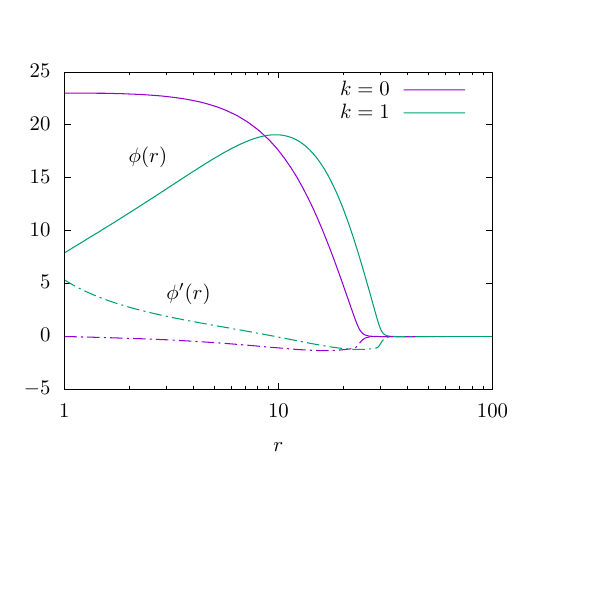}
\end{center}
\vspace{-2cm}

\caption{{\it Left}: We show the scalar field function $\phi$ for clouds on electrically and magnetically charged RN black holes with $r_h=1.0$, $Q_e=0.12$, $n=1$ and three different values of the gauge coupling $g$. Note that different values of $g$ correspond to different magnetic charges $Q_m$ of the RN solutions, $Q_m=1/(2g)$.  
{\it Right}: We show the profile of $\phi(r)$ as well as of $\phi'(r)$ for scalar clouds on RN black holes with $r_h=1.0$, $Q_e=0.12$, $g=1.0$ and two different values of $k$. Note that for $k=0$ the RN black hole is only electrically charged, while for $k=1$, the 
RN black hole possesses additionally a magnetic charge $Q_m=0.5.$
 \label{fig:profile_dyon_probe}
}
\end{figure}

%%%%%%%%%%%%%%%%%%%%%%%%%%%%%%%%%%%%%%%%%%%%%
We now turn to the case $k=1$. We show the scalar field function $\phi(r)$ for clouds
on electrically and magnetically charged RN black holes with $r_h=1.0$ and $Q_e=0.12$ 
in Fig. \ref{fig:profile_dyon_probe} (left).
A general feature of these solutions is that the scalar field develops a local maximum at some intermediate radius. We find that increasing $g$ and hence decreasing the magnetic charge $Q_m$ of the background RN solution that the value of the local maximum of $\phi(r)$ decreases and is located at decreased values of the radial coordinate $r$. To understand the difference to the $k=0$ case, we show the profiles of $\phi(r)$ and $\phi'(r)$ in Fig. \ref{fig:profile_dyon_probe} (right) for $r_h=1.0$, $Q_e=0.12$, $g=1.0$ and compare the cases $k=0$ and $k=1$. Clearly, for the same value of the horizon radius and the same value of the gauge coupling $g$, scalar clouds on electrically charge RN black holes
are qualitatively different to scalar clouds on electrically and magnetically charged RN black holes: the scalar field has larger values on the horizon and the clouds extend to smaller radii in the former case. The presence of the magnetic charge seems to push the
scalar cloud away from the horizon.

%%%%%%%%%%%%%%%%%%%%%%%%%%%%%%%%%%%%%%%%%%%%
%%%%%%%%%%%%%%%%%%%%%%%%%%%%%%%%%%%%%%%%%%%

\begin{figure}[h!]
\begin{center}
\includegraphics[width=8cm]{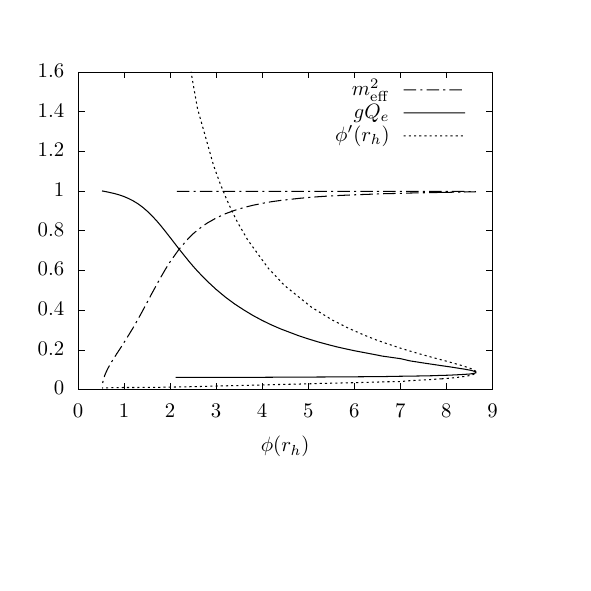}
\includegraphics[width=8cm]{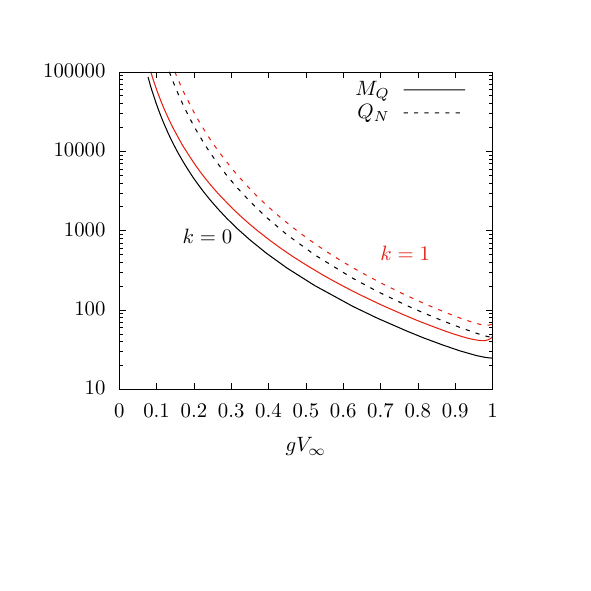}
\end{center}
\vspace{-2cm}

\caption{{\it Left}~: We show the dependence of $m_{\rm eff}^2$ (dotted-dashed), $gQ_e$ (solid) and $\phi'(r_h)$ (dotted) on $\phi(r_h)$ for scalar clouds on RN black holes with $r_h=1$, $Q_e= 0.12$ and $k=1$. 
{\it Right}~: We show the dependence of the mass $M$ (solid) and the Noether charge $Q_N$ (dashed) on $gV_{\infty}$ for scalar clouds on RN black holes with $r_h=1$ and $Q_e= 0.12$. We compare the
$k=0$ (black) and $k=1$ (red) case.
 \label{fig:data_probe_dyon}
}
\end{figure}

Our numerical results also confirm the validity of (\ref{eq:g_condition}), i.e. the fact that for scalar clouds to exist on electrically and magnetically charged RN black holes
$g$ cannot be arbitrarily small. The interval of $g$ for which non-trivial scalar field solutions exist, becomes smaller when either decreasing $r_h$ or increasing $Q_e$. 
In Fig. \ref{fig:data_probe_dyon} we show the dependence of $gQ_e$, $m_{\rm eff}^2$, $\phi'(r_h)$ on $\phi(r_h)$ (left) as well as the dependence of $M_Q$ and $Q_N$ on $g V_{\infty}$ (right) for $r_h=1.0$ and $Q_e=0.12$. The comparison of the $k=0$ and $k=1$ demonstrates that while we find only one branch of solutions in $\phi(r_h)$ when $k=0$, we find
two branches when $k=1$. Moreover, these two branches join at a maximal value of $\phi(r_h)$, while $\phi(r_h)$ can tend to infinity for $k=0$. One branch terminates at
$m_{\rm eff}^2 =0$ (similar to the $k=0$ case), the second branch bifurcates with the 
extremal RN solution. Our numerical results confirm that this limit is reached for $g\approx 0.503$ which is in perfect agreement with (\ref{eq:g_condition}). Extremal RN black holes cannot carry regular scalar hair and hence we find that 
$\phi'(r_h)\rightarrow \infty$ in this limit. In Fig. \ref{fig:data_probe_dyon} (right) we show the dependence of the mass $M_Q$ and the Noether charge $Q_N$ on $gV_{\infty}$. We find only one branch of solutions
for both $k=0$ and $k=1$ and that both the mass $M_Q$ and the Noether charge $Q_N$ tend to finite
values in the limit $gV_{\infty}\rightarrow 1$. On the other hand, $M_Q$ and $Q_N$ diverge for $gV_{\infty}\rightarrow 0$. The clouds on electrically charged RN black holes have lower values of $M_Q$ and $Q_N$ than their counterparts on electrically and magnetically charged RN black holes. For both $k=0$ and $k=1$ we find that the Noether charge is larger than the mass $M_Q$.
%%%%%%%%%%%%%%%%%%%%%%%%%%%%%%%%%%%%%%%%%%%%
%%%%%%%%%%%%%%%%%%%%%%%%%%%%%%%%%%%%%%%%%%%%
%%%%%%%%%%%%%%%%%%%

%%%%%%%%%%%%%%%%%%%%%%%%%%%%%%%%%%%%%%%%%%%%%%%%%%%%%%%%%%%%%%%%%%%%%%%%%%%%%%%
\subsection{Charged scalar clouds on Schwarzschild black holes}
%%%%%%%%%%%%%%%%%%%%%%%%%%%%%%%%%%%%%%%%%%%%%%%%%%%%%%%%%%%%%%%%%%%%%%%%%%%%
Another interesting  limiting case is $\alpha=0$, i.e. the limit in which all matter fields
decouple from gravity. The unique solution to the vacuum Einstein equation is the Schwarzschild space-time with
 $N(r) = 1 - r_h/r$ , $\sigma(r) \equiv 1$. This solution is uniquely characterized by the event horizon $r_h$.

%%%%%%%%%%%%%%%%%%%%%%%%%%%%%%%%%%%%%%%%%%%%%%%%%%%%%%%%%%%%%%%%%%%%
\subsection{Electrically charged clouds on Schwarzschild black holes}
%%%%%%%%%%%%%%%%%%%%%%%%%%%%%%%%%%%%%%%%%%%%%%%%%%%%%%%%%%%%%%%%%%%%%

Let us first discuss the case $k=0$. We find that non-trivial scalar fields
exist for  $g \in ] 0,  g_{\rm max}]$ and that $g_{\rm max}$ depends on the event horizon radius $r_h$, e.g. $g_{\rm max}\approx 0.11$ for $r_h=0.15$, while $g_{\rm max}\approx 0.05$
for $r_h=1.0$, i.e. larger (and more massive) Schwarzschild black holes require smaller values of the gauge coupling $g$.  The solutions can be characterized uniquely by the value of the electric field on the horizon $\sim V'(r_h)$ and exist for $V'(r_h) \in [W_1,W_2]$
where $W_{1,2}$ depend on $g$ and $r_h$. For $g \to g_{\rm max}$ we find that $W_1\rightarrow W_2$. In the two limits $V'(r_h) \to W_{1,2}$ the effective mass of the scalar field, $m^2-g^2 V_{\infty}^2$, tends to zero, i.e. the scalar field is no longer exponentially localized.  In the Table \ref{table:gW1W2} we give the values of the minimal and maximal value of the electric field on the horizon for $r_h=0.15$ and different values of $g$. Obviously, charged scalar clouds exist on Schwarzschild black holes only when the electric field is sufficiently large and at the same time not too large.
Clearly, the electric field is necessary to allow for scalar clouds to exist on Schwarzschild black holes.

\begin{table}[h!]
\centering
 \begin{tabular}{|c| c |c|} 
 \hline
 $g$ & $W_1$ & $W_2$  \\ [0.5ex] 
 \hline\hline
 $0.01$ & $0.8$ & $660.0$ \\
  $0.02$ & $4.2$ & $330.0$ \\
 $0.08$ & $22.0$ & $78.0$ \\
  $0.10$ & $27.0$ & $59.0$ \\
   $0.11$ & $45.0$ & $45.0$ \\
   \hline
 \end{tabular}
 \caption{The minimal ($W_1$) and maximal ($W_2$) value of the electric field $V'(r_h)$ of a scalar cloud on the horizon of a Schwarzschild black hole with $r_h=0.15$. \label{table:gW1W2}}
\end{table}

\begin{figure}[h!]
\begin{center}
\includegraphics[width=8cm]{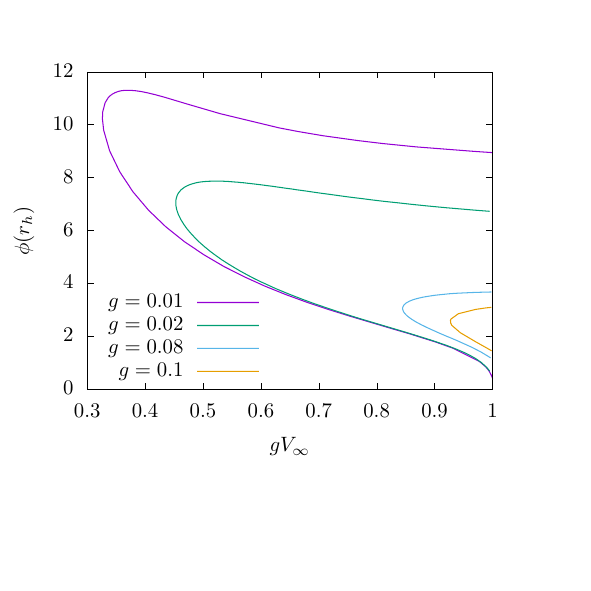}
\includegraphics[width=8cm]{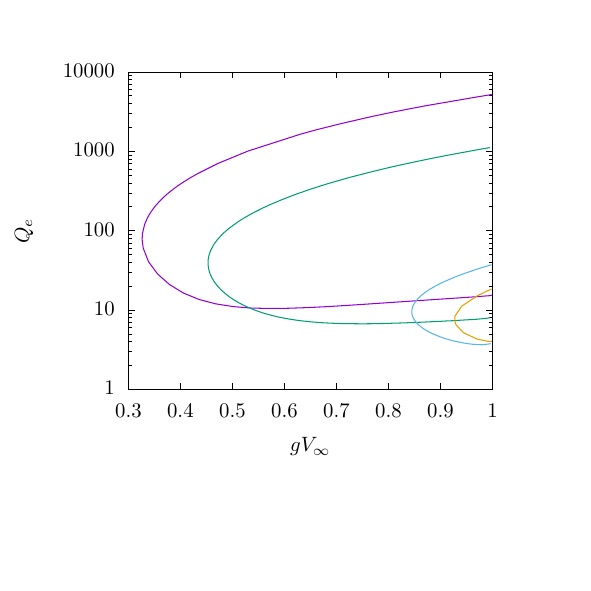}
\end{center}
\vspace{-2cm}
\caption{{\it Left}~: We show the dependence of the scalar field on the horizon, $\phi(r_h)$, on
$gV_{\infty}$ for electric clouds on Schwarzschild black holes with $r_h=0.15$ and several values of $g$. 
{\it Right}~: Same as left, but for the electric charge $Q_e$ of the scalar cloud.
 \label{fig:electric_schwarzschild}
}
\end{figure}

In Fig. \ref{fig:electric_schwarzschild} we 
give the value of the scalar field on the horizon, $\phi(r_h)$, (left) and the electric charge 
$Q_e$ (right) of the scalar cloud in dependence of $gV_{\infty}$. Clearly, who branches appear in $gV_{\infty}$ which both end at $gV_{\infty}=1$ and merge at a minimal value of $gV_{\infty}$. This minimal value of $gV_{\infty}$ corresponds to an intermediate value of the electric field on the horizon, $V^{\prime}(r_h)$. 
The lower branch in $\phi(r_h)$ corresponds to larger values of the electric field on the horizon, $V^{\prime}(r_h)$, while the upper
branch has lower values of the electric field on the horizon. Similarly, 
the lower branch in $Q_e$ corresponds to larger values of $V^{\prime}(r_h)$, while the upper branch corresponds to smaller
values of $V^{\prime}(r_h)$. Hence, electrically charged scalar clouds on Schwarzschild black holes
have either small values of the scalar field on the horizon with large electric fields and small electric charge of the cloud, or they have large scalar fields with small electric fields
and large electric charge of the cloud.

%%%%%%%%%%%%%%%%%%%%%%%%%%%%%%%%%%%%%%%%%%%%
%%%%%%%%%%%%%%%%%%%

%%%%%%%%%%%%%%%%%%%%%%%%%%%%%%%%%%%%%%%%%%%%%%%%%%%%%%%%%%%%%%%
\begin{figure}[h!]
\begin{center}
\includegraphics[width=8cm]{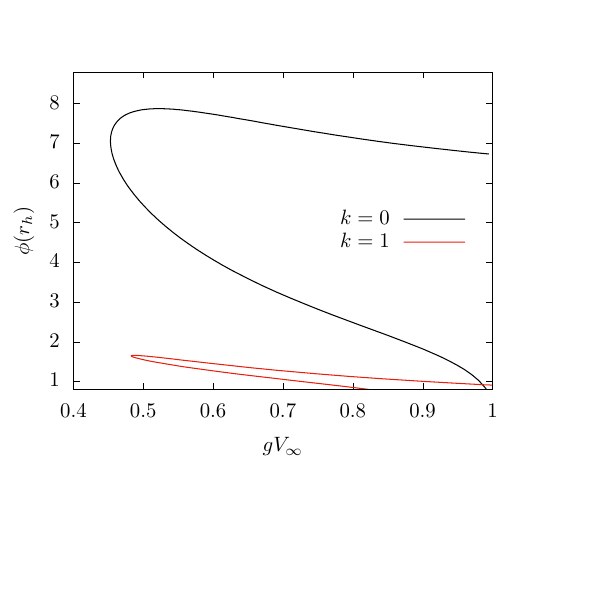}
\includegraphics[width=8cm]{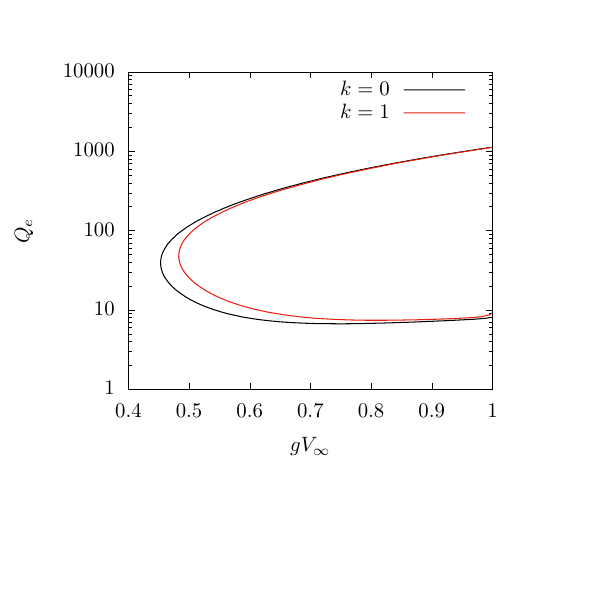}
\end{center}
\vspace{-2cm}
\caption{{\it Left}~: We show the value of the scalar field on the horizon $\phi(r_h)$ in dependence of $gV_{\infty}$
for electric ($k=0$, black) and dyonic ($k=1$, red) clouds on Schwarzschild black holes with $r_h=0.15$ and $g=0.02$.
{\it Right}~: Same as left, but for the electric charge of the scalar cloud $Q_e$.
\label{fig:Qe_vs_gVh}
}
\end{figure}
%%%%%%%%%%%%%%%%%%%%%%%%%%%%%%%%%%%%%%%%%%%%%%%%%%%%%%%%%%%%%%%%

%%%%%%%%%%%%%%%%%%%%%%%%%%%%%%%%%%%%%%%%%%%%%%%%%%%%%%%%%%%%%%%%%%%%%%%%%%
\subsubsection{Electrically and magnetically charged clouds on Schwarzschild black holes}
%%%%%%%%%%%%%%%%%%%%%%%%%%%%%%%%%%%%%%%%%%%%%%%%%%%%%%%%%%%%%%%%%%%%%%%%%%
Setting  $k=1$, we study electrically and magnetically charged scalar clouds on Schwarzschild black holes. In order to understand the influence of the magnetic charge on the solutions, we compare the cases $k=0$ and $k=1$ in Fig. \ref{fig:Qe_vs_gVh}, where we give the scalar field on the horizon $\phi(r_h)$ (left) and the electric charge of the scalar cloud $Q_e$ (right) in dependence of $gV_{\infty}$ 
for $g=0.02$ and $r_h=0.15$. Again, we find two branches of solutions in $g V_{\infty}$ which both end at $g V_{\infty}=1$ and join at a minimal value of $gV_{\infty}$. This minimal value is slightly smaller for $k=0$ solutions
as compared to $k=1$ solutions.  While the additional magnetic charge does not change the value of $Q_e$ strongly, the value of the scalar field on the horizon $\phi(r_h)$ is much smaller for $k=1$ as compared to $k=0$ solutions. Electrically charged scalar fields can possess much larger values on the event horizon of a Schwarzschild black hole as electrically and magnetically charged scalar fields.

%%%%%%%%%%%%%%%%%%%%%%%%%%%%%%%%%%%%%%%%%%%%%%%%%%%%%%%%%%%%%%%%%%%
%%%%%%%%%%%%%%%%%%%%%%%%%%%%%%%%%%%%%%%%%%%%%%%%%%%%%%%%%%%%%%%%%%%
%%%%%%%%%%%%%%%%%%%%%%%%%%%%%%%%%%%%%%%%%%%%%%%%%%%%%%%%%%%%%%%%%%%
\section{Charged black holes with scalar hair}
\begin{figure}[h!]
\begin{center}
\includegraphics[width=8cm]{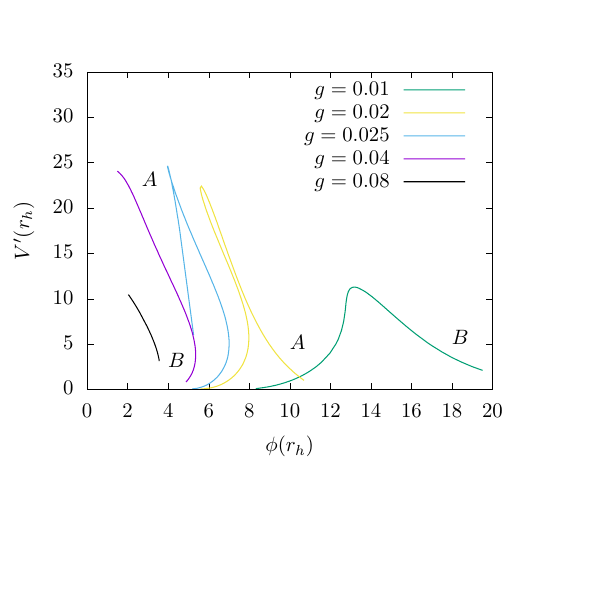}
\includegraphics[width=8cm]{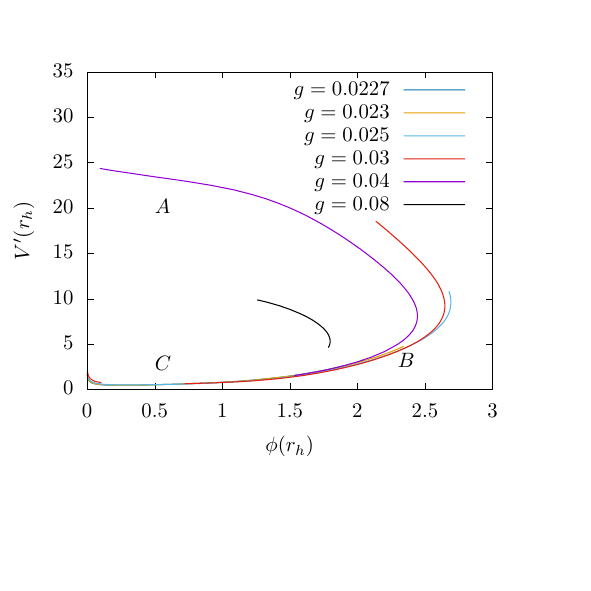}
\end{center}
\vspace{-2cm}
\caption{{\it Left}~: We show the dependence of the value of the electric field on the horizon, $V^{\prime}(r_h)$, on the value of the scalar field on the horizon, $\phi(r_h)$ for electrically charged black holes ($k=0$) with scalar hair for $\alpha=0.01$, $r_h=1.0$ and several values of the gauge coupling $g$. {\it Right}~: Same as left, but for electrically and magnetically charged, i.e. dyonic black holes ($k=1$). 
\label{fig:vp_phi_relation}
}
\end{figure}

%%%%%%%%%%%%%%%%%%%%%%%%%%%%%%%%%%%%%%%%%%%%%%%%%%%%%%
In the fully backreacted case, we are left with three free parameters after appropriate rescaling of the coordinate and fields, respectively. These are the gravitational coupling
$\alpha$, the gauge coupling $g$ and the radius of the event horizon of the black hole $r_h$.  We find that for a fixed choice of $\alpha$, $g$ and $r_h$  a family  of solutions can be constructed by using the value of the electric field on the horizon, $V'(r_h)$, or, alternatively, the value of the scalar field on the horizon, $\phi(r_h)$. We find several branches in these parameters. First, let us give the relation between these two parameters. This is shown in Fig. \ref{fig:vp_phi_relation}, where we compare the purely electric case (left, $k=0$) with the dyonic case (right, $k=1$). Clearly, the pattern is qualitatively very different for the two cases, in particular, the value of the scalar field on the horizon, $\phi(r_h)$ can be much larger for electrically charged black holes as compared to dyonic black holes.

\begin{figure}[h!]
\begin{center}
\includegraphics[width=8cm]{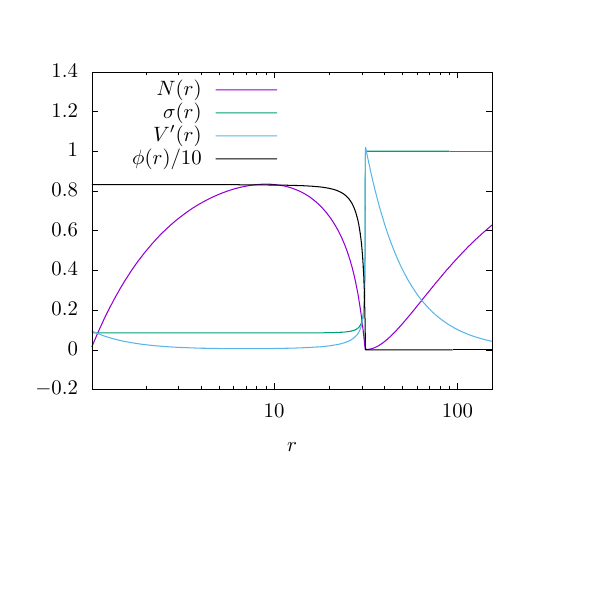}
\includegraphics[width=8cm]{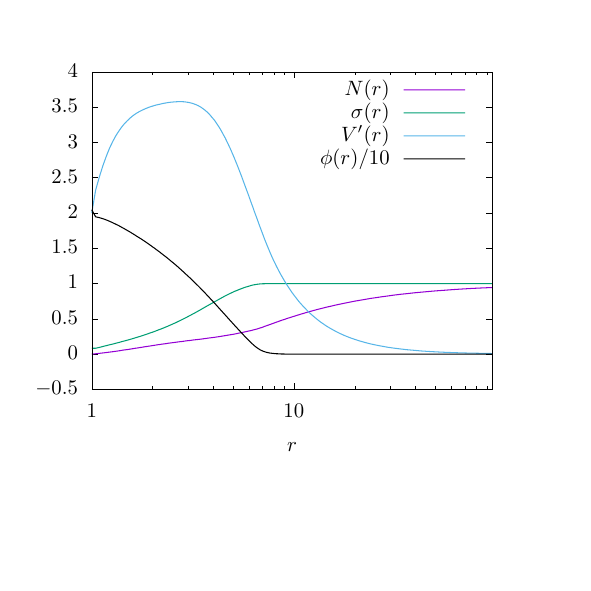}
\end{center}
\vspace{-2cm}
\caption{{\it Left}~: We show the profile of the metric function $N$, of the electric field  $V^{\prime}$, and the scalar field $\phi$ for the solution on branch $A$ (see Fig.\ref{fig:vp_phi_relation}) with  $g=0.01$ and
$V'(r_h)=0.1$.
{\it Right}~: Same as left, but for the solution on branch $B$ with $g=0.01$
and $V'(r_h)=2.0$.  
\label{fig:profile_EBH}
}
\end{figure}

In the $k=0$ case, we find that for sufficiently small
values of $g$ two branches exist, labeled $A$ and $B$ in Fig. \ref{fig:vp_phi_relation} (left). The solutions on these two branches are very different. This is shown in Fig. \ref{fig:profile_EBH}, where we give the profiles of the metric function $N(r)$ and $\sigma(r)$ as well as the electric field $V^{\prime}(r)$ and the scalar field $\phi(r)$, respectively, for a solution on the branch $A$ (left) and for a solution on the branch $B$ (right) for $g=0.01$ and $r_h=1.0$. The solution from branch $A$ has a small electric field on the horizon, $V^{\prime}(r_h)=0.1$, and clearly shows a hard wall at $r_c > r_h$ such that for $r > r_c$ the solution
corresponds to an extremal RN solution with vanishing scalar field $\phi(r)\equiv 0$, while for $r < r_c$
the scalar field is constant, but non-vanishing. As was discussed previously in \cite{bh1,bh2,bh3}, the interior of this solution can be interpreted as inflating with the constant scalar field energy equal to the positive cosmological constant. On the other hand, the solution on branch $B$ has a large electric field on the horizon, $V^{\prime}(r_h)=2.0$. The scalar field becomes identically zero for large enough $r$ at $r=r_c$ such that for $r > r_c$ the solution is given by $\phi\equiv 0$, $\sigma\equiv 1$, while $N(r)$ and $V'(r)$ are non-trivial and, in particular, $N(r_c) > 0$.

\begin{figure}[h!]
\begin{center}
\includegraphics[width=8cm]{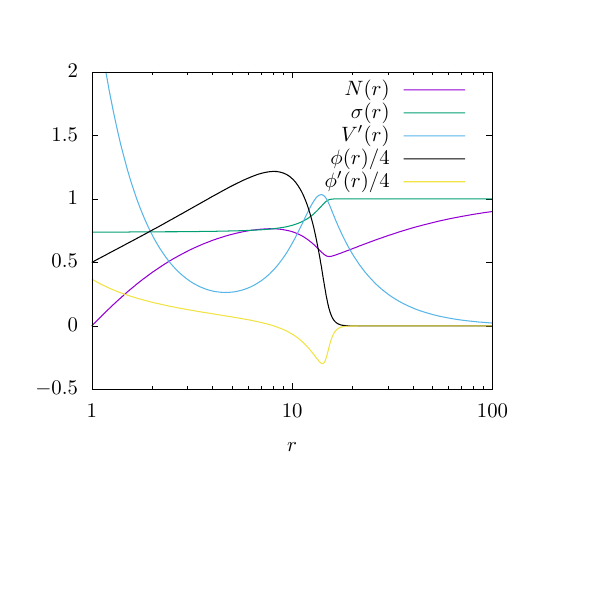}
\includegraphics[width=8cm]{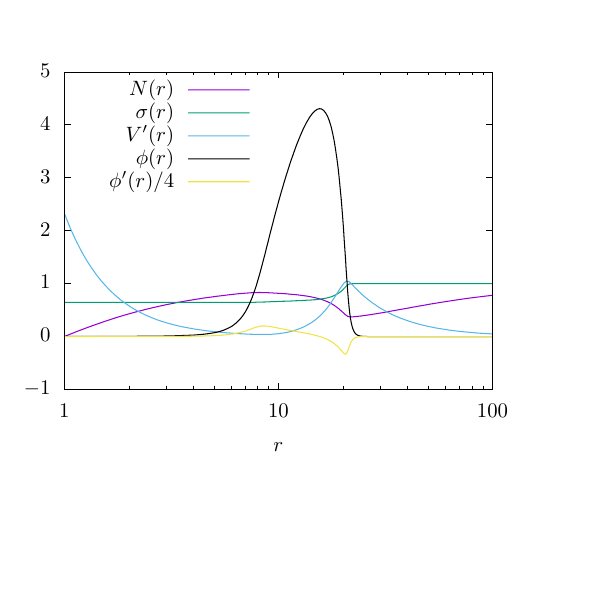}
\end{center}
\vspace{-2cm}
\caption{{\it Left:} We show the profiles of the metric functions $N(r)$  and $\sigma(r)$, the electric field $V^{\prime}(r)$, the scalar field $\phi(r)$ and the scalar field derivative $\phi^{\prime}(r)$ for a dyonic black hole with $r_h=1.0$, $\alpha=0.001$, $g=0.03$ and $\phi(r_h)=2.0$.  {\it Right:}  We show the profiles of the metric functions $N(r)$ and $\sigma(r)$, the electric field $V^{\prime}(r)$, the scalar field $\phi(r)$ and the scalar field derivative $\phi^{\prime}(r)$ for a dyonic black hole with $r_h=1.0$, $\alpha=0.001$, $g=0.03$ and $\phi(r_h)=0.0002$. 
\label{fig:dyonic_solutions}
}
\end{figure}

For the $k=1$ solutions, we find a third branch (labeled $C$ in Fig. \ref{fig:vp_phi_relation}) for intermediate values of $g$. In order to understand this new branch, we show the profiles of 
solutions for $r_h=1.0$, $\alpha=0.001$, $g=0.03$ in Fig. \ref{fig:dyonic_solutions}. On the left
of this figure a solution for $\phi(r_h)=2.0$ which corresponds to the merging point of branches $A$ and $B$ in Fig. \ref{fig:vp_phi_relation} (right, red line) is shown. The scalar field becomes identically zero for $r > r_c$ and the electric field corresponds to that of RN solution. For $r_h < r < r_c$, the solution is non-trivial and, in particular, the scalar field does no longer have its maximal value at $r_h$. The scalar cloud is pushed away from the horizon when adding magnetic charge. 
On the right of Fig. \ref{fig:dyonic_solutions} we show a solution on branch $C$. This solution has very small scalar field and $\sigma=\sigma_0={\rm constant} < 1$ on and outside the horizon up to some intermediate value of $r=r_0$. For $r > r_0$ and $r< r_c$, the scalar field (and with it the remaining matter and metric functions) are non-trivial, while for $r > r_c$, we find $\phi(r)\equiv 0$, $\sigma\equiv 1$, the electric field becomes equal to that of a RN solution, $V'=V'_{RN}$, while $N(r)$ possesses a minimum. Our numerical results suggest that for the values of coupling constants chosen here, the minimum actually never drops down to zero. 
Hence, in contrast to the electric case, we do not find ``hard wall'' solutions with an inflating interior, but rather a solution that has different behaviour on the intervals $r\in [r_h,r_0]$, $[r_0,r_c]$ and $[r_c,\infty[$. The solution for $[r_c,\infty[$ is an extremal RN solution, while the solution in $[r_0,r_c]$ is non-trivial. The solution for $r\in [r_h,r_0]$ is given by
\begin{equation}
V(r)=V_{\infty} + \frac{Q_e}{r} \ \ , \ \ N(r)=1-\frac{2M_{\rm ADM}}{r} + \frac{\alpha Q_e^2}{\sigma_0^2 r^2} + \frac{\alpha k^2}{4g^2 r^2} \ \ .
\end{equation}
Note that the electric charge is modified by a factor of $1/\sigma_0$ in $N(r)$.

\begin{figure}[h!]
\begin{center}
\includegraphics[width=8cm]{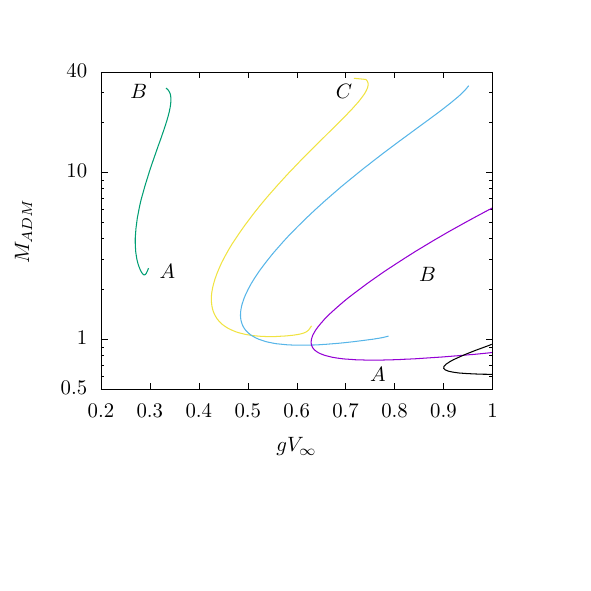}
\includegraphics[width=8cm]{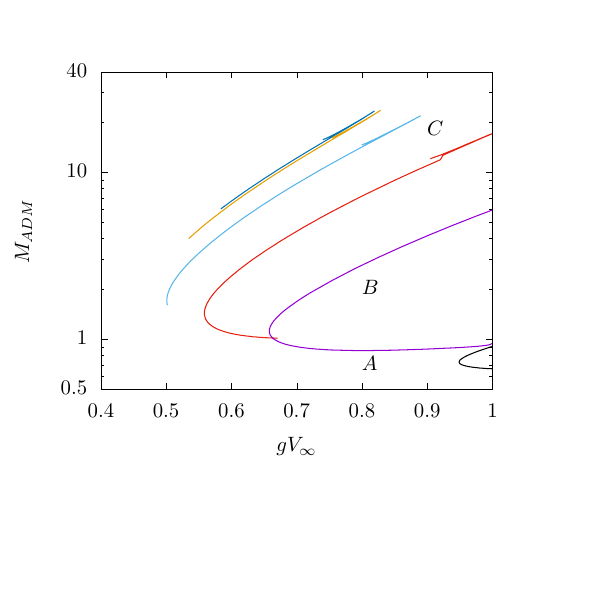}
\end{center}
\vspace{-2cm}
\caption{We show the dependence of the ADM mass $M_{ADM}$ on $gV_{\infty}$ for $\alpha=0.01$, $r_h=1.0$ and several values of $g$ - colour-coding as in Fig. \ref{fig:vp_phi_relation}. We compare the purely electric case $k=0$ (left) with the dyonic case $k=1$ (right). 
\label{fig:M_gvh_compare}
}
\end{figure}

In Fig. \ref{fig:M_gvh_compare}, we show the ADM mass of the solutions, $M_{ADM}$, in dependence of $gV_{\infty}$ for $\alpha=0.01$, $r_h=1.0$ and several values of $g$. We compare the $k=0$ (left) and the $k=1$ (right) case. We find that for large values of $g$, the electrically charged and the dyonic black holes show the same pattern~: there exist two branches of solutions which both terminate at $gV_{\infty}=1$. These branches are labeled $A$ and $B$ in the plot. 

\begin{figure}[h!]
\begin{center}
\includegraphics[width=8cm]{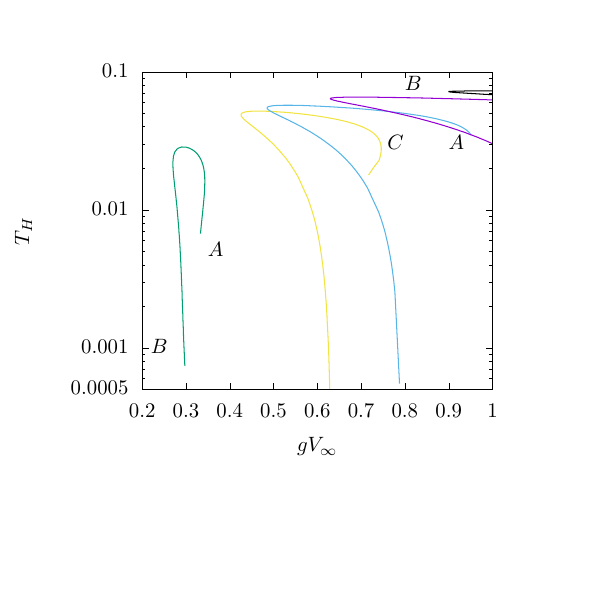}
\includegraphics[width=8cm]{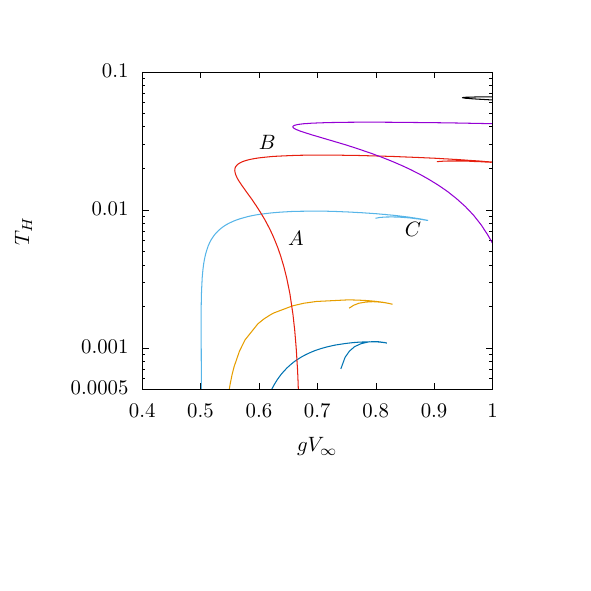}
\end{center}
\vspace{-2cm}
\caption{We show the dependence of the Hawking temperature $T_H$ on $gV_{\infty}$ for $\alpha=0.01$, $r_h=1.0$ and several values of $g$ - colour-coding as in Fig. \ref{fig:vp_phi_relation}. We compare the purely electric case $k=0$ (left) with the dyonic case $k=1$ (right). 
\label{fig:TH_gvh_compare}
}
\end{figure}

For lower values of the gauge coupling $g$ branch $C$ appears here as well. Interestingly, some of these branches terminate at intermediate values of $gV_{\infty}$, i.e. at $gV_{\infty} < 1$. Our numerical results indicate that the qualitative features of these new branches differ when comparing $k=0$ and $k=1$. To understand these features, it is instructive to compare the values of the Hawking temperature $T_H$. This is shown in Fig. \ref{fig:TH_gvh_compare}.  We find that for $g > 0.04$, all solutions have non-zero temperature and 
the solutions with lowest ADM mass have also the lowest temperature $T_H$. 
However,  our numerical results strongly suggest that for $g < 0.04$ we find solutions with $T_H \to 0$. However, due to limited numerical accuracy, we can only rely on our results for $T_H \gtrsim 10^{-3}$. 

\begin{figure}[h!]
\begin{center}
\includegraphics[width=8cm]{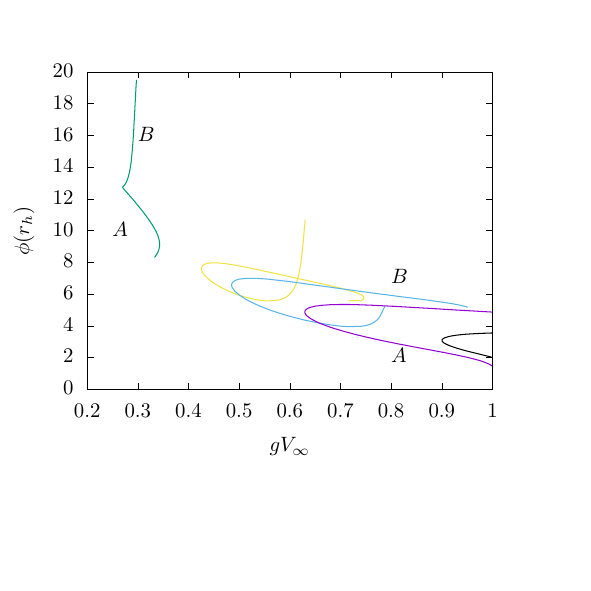}
\includegraphics[width=8cm]{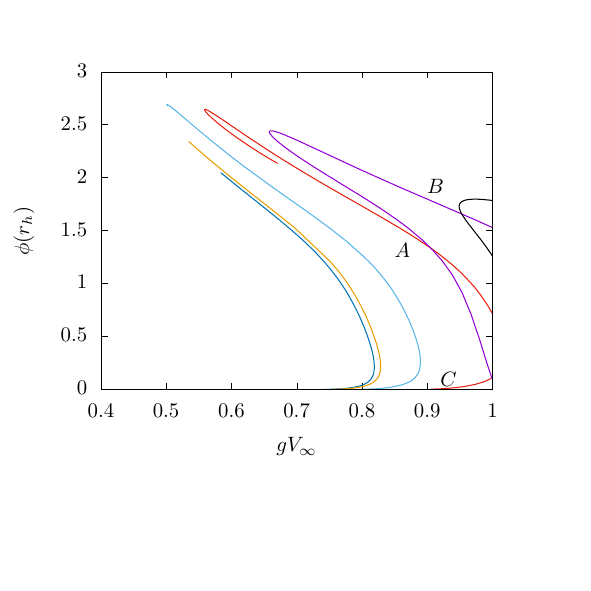}
\end{center}
\vspace{-2cm}
\caption{We show the dependence of the value of the scalar field on the horizon, $\phi(r_h)$, on $gV_{\infty}$ for $\alpha=0.01$, $r_h=1.0$ and several values of $g$ - colour-coding as in Fig. \ref{fig:vp_phi_relation}. We compare the purely electric case $k=0$ (left) with the dyonic case $k=1$ (right). 
\label{fig:phirh_gvh_compare}
}
\end{figure}

\begin{figure}[h!]
\begin{center}
\includegraphics[width=8cm]{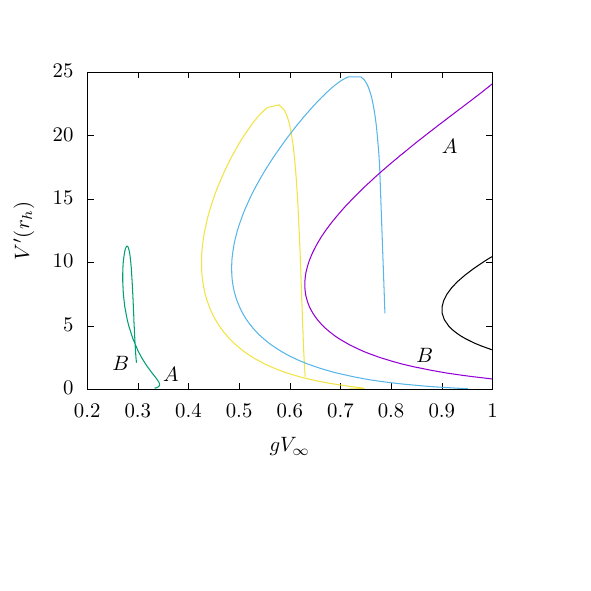}
\includegraphics[width=8cm]{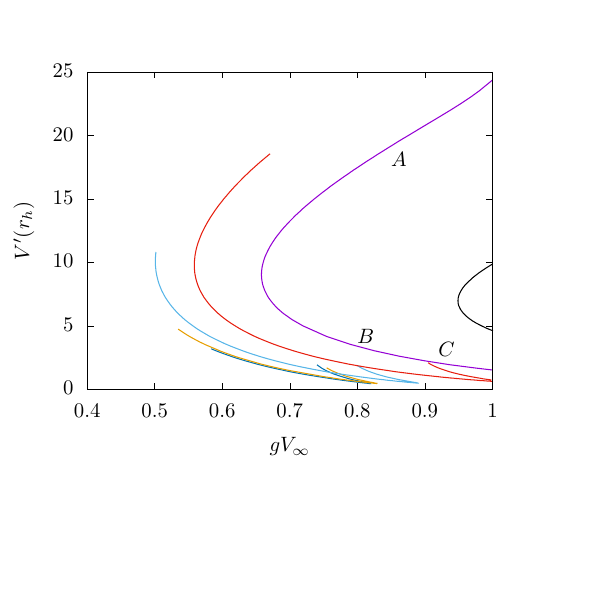}
\end{center}
\vspace{-2cm}
\caption{We show the dependence of the value of the electric field on the horizon, $V^{\prime}(r_h)$, on $gV_{\infty}$ for $\alpha=0.01$, $r_h=1.0$ and several values of $g$ - colour-coding as in Fig. \ref{fig:vp_phi_relation}. We compare the purely electric case $k=0$ (left) with the dyonic case $k=1$ (right). 
\label{fig:Vprh_gvh_compare}
}
\end{figure}

In Fig. \ref{fig:phirh_gvh_compare} we show the value of the scalar field on the horizon, $\phi(r_h)$, in function of $g V_{\infty}$ for $\alpha=0.01$ and $r_h=1.0$ and several values of $g$. We note that the presence of the magnetic charge allows for solutions with $\phi(r_h)\approx 0$. In the electric case, $\phi(r_h)=0$ would imply $\phi(r)\equiv 0$, however, since the presence of the magnetic charge leads to a maximal value of $\phi(r)$ away from the horizon $\phi(r_h)\approx 0$ does not imply a trivial scalar field in the dyonic case. Close to $\phi(r_h)=0$, we observe the existence of branch C. This branch can also be seen in Fig. \ref{fig:Vprh_gvh_compare}. It corresponds not only to very small scalar fields at $r_h$, but also to increased electric fields on the horizon and decreased potential difference between the horizon and infinity when compared to branch B.

\begin{figure}[h!]
\begin{center}
\includegraphics[width=8cm]{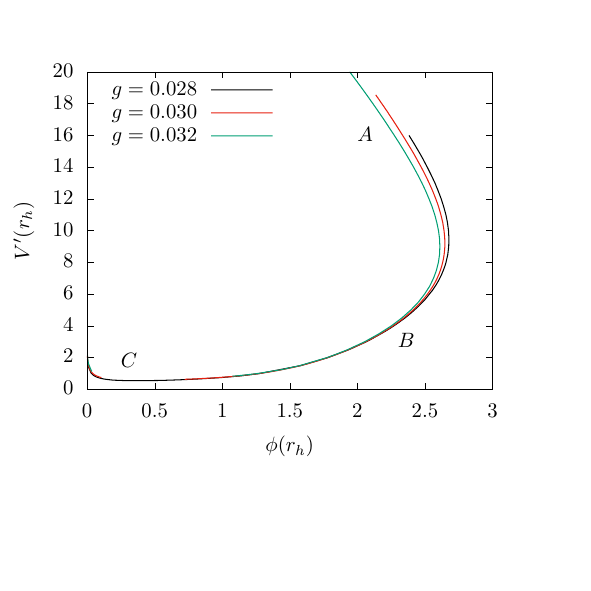}
\includegraphics[width=8cm]{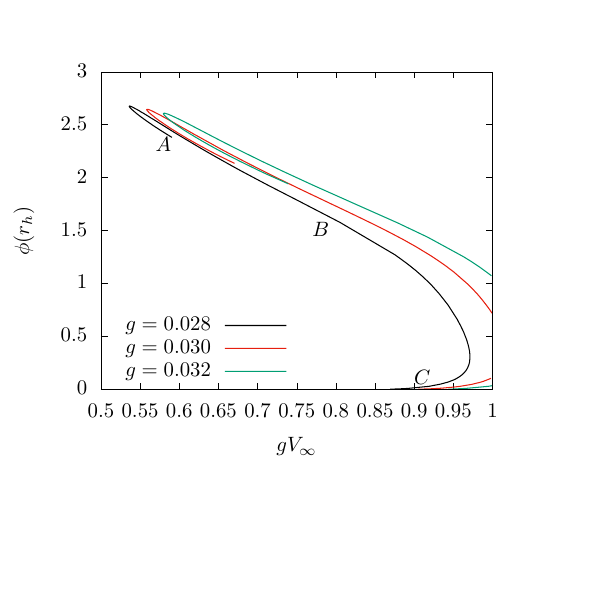}
\end{center}
\vspace{-2cm}
\caption{{\it Left:} We show the dependence of the value of the electric field on the horizon, $V^{\prime}(r_h)$, on the value of the scalar field on the horizon, $\phi(r_h)$, for $\alpha=0.01$, $r_h=1.0$ and three values of $g$ close to $g=0.03$. {\it Right:} We show the dependence of the value of the value of the scalar field on the horizon, $\phi(r_h)$, on $gV_{\infty}$ for $\alpha=0.01$, $r_h=1.0$ and three values of $g$ close to $g=0.03$. 
\label{fig:branchC}
}
\end{figure}

Let us finally note that the emergence of branch $C$ for $g\approx 0.03$ came as a surprise. 
To understand this branch $C$ better, we show the dependence of the value of the electric field on the horizon, $V^{\prime}(r_h)$, on the value of the scalar field on the horizon, $\phi(r_h)$, for $\alpha=0.01$, $r_h=1.0$ and three values of $g$ close to $g=0.03$ in Fig. \ref{fig:branchC} (left). The new branch appears for sufficiently small values of $\phi(r_h)$ and corresponds to small values of the electric field on the horizon.
The dependence of $\phi(r_h)$ on $gV_{\infty}$ show in the same figure on the right demonstrates that the limitation of $gV_{\infty} \leq 1$ leads to the emergence of this new branch. While for $g=0.028$ and $g=0.03$ the branches A, B and C combine into one smooth curve, this is no longer true for $g=0.032$, where the smooth curve is split into two at $gV_{\infty}=1$. Hence for $g=0.032$ branch C is disconnected from the branches A and B and a gap of values in $\phi(r_h)$ appears such that for these values no dyonic black holes with scalar hair exist.

\section{Conclusions}
In this paper, we have studied static, spherically symmetric black hole solutions that possess non-trivial scalar hair. These solutions exist in 
a complex scalar field model when coupled minimally to a U(1) gauge field.
Next to electrically charged black holes, we have also discussed black holes
with electric and magnetic charge. This is possible in a U(1) gauge theory
since the source of the magnetic field is hidden behind the event horizon of the black hole.
In comparison to a previous study done in \cite{Herdeiro:2024yqa}, we have
used a scalar field potential that is exponential in nature and hence bounded.
Moreover, we have put the emphasis on studying the influence of the gauge coupling constant $g$, in particular for the cases of fixed background black hole solutions. We have also tried to directly compare the electric and the dyonic case. We find that the presence of the magnetic charge leads to some interesting qualitative features. Dyonic Reissner-Nordstr\"om black holes
cannot carry arbitrarily large values of the scalar field on the horizon,  while there is no upper limit on $\phi(r_h)$ for electrically charged Reissner-Nordstr\"om black holes. Moreover, we find that dyonic black holes
can carry scalar fields that are actually very small on the horizon, something that is not possible for the purely electrically charged black holes. \\
\\{\bf Acknowledgments} Katherine Horton is supported by the Engineering and Physical Sciences Research Council EP/W524335/1.

\clearpage

%%%%%%%%%%%%%%%%%%%%%%%%%%%%%%%%%%%%%%%%%%%%%%%%%%%%%%%%%%%

\end{document}